\documentclass[preprint2]{aastex}

\newcommand{\mjup}{M$_{\mbox{\footnotesize Jup}}$}
\newcommand{\muas}{$\mu$as}

\newcommand{\ms}{m\,s$^{-1}$}
\newcommand{\msun}{M$_{\odot}$}

\slugcomment{Based on observations collected at Lick Observatory}

\shorttitle{Pollux and its Planetary Companion}
\shortauthors{Reffert et al.}

\begin{document}


\title{Precise Radial Velocities of Giant Stars\\
II. Pollux and its Planetary Companion}

\author{Sabine Reffert and Andreas Quirrenbach}
\affil{ZAH - Landessternwarte, K\"onigstuhl~12, 69117~Heidelberg, Germany}
\email{sreffert@lsw.uni-heidelberg.de}

\author{David S.~Mitchell}
\affil{California Polytechnic State University, San Luis Obispo, CA 93407, USA}

\author{Simon Albrecht and Saskia Hekker}
\affil{Sterrewacht Leiden, Niels Bohrweg 2, 2333 CA Leiden, The Netherlands}

\author{Debra A.~Fischer}
\affil{Department of Physics and Astronomy, San Francisco State University,
CA 94132, USA}

\author{Geoffrey W.~Marcy}
\affil{Department of Astronomy, University of California, Berkeley, CA~94720,
USA}

\and

\author{R.~Paul Butler}
\affil{Department of Terrestrial Magnetism, Carnegie Institution of
Washington, 5241 Broad Branch Road NW, Washington, DC~20015, USA}



\begin{abstract}
It has long been speculated that the observed periodic radial velocity pattern
for the K~giant Pollux might be explained in terms of an orbiting planetary
companion. We have collected 80 high-resolution spectra for Pollux at
Lick Observatory yielding precise radial velocities with a mean error of
3.8~m/s, providing the most comprehensive and precise data set available
for this star. Our data confirm the periodicity previously seen in the radial 
velocities. We derive a period of 589.7$\pm$3.5~days and, assuming a
primary mass of 1.86~\msun, a minimum companion mass of 
2.9$\pm$0.3\mjup, consistent with earlier determinations.
No evidence for any periodicities is visible in our analysis of
the shapes of the spectral lines via the bisector method, so that we
conclude that evidence is accumulating and compelling for a planet around
Pollux. However, some last doubt remains about this interpretation, because 
non-radial pulsations which might be present in giant stars could in principle
also explain the observed radial velocities, while the accompanying bisector
variations might be too small to be detectable with current data.
\end{abstract}



\section{Introduction}

Pollux ($\beta$~Gem, HR~2990, HD~62509, HIP~37826) is one of the brightest
stars in the sky (V=1.16~mag) and has been observed extensively in the past.
Fundamental parameters from a detailed model atmosphere analysis of the 
spectrum have e.g.\ been provided by \citet{drake91}, and it is usually 
classified as K0IIIb star \citep{keenan89}. The parallax determined by
Hipparcos results in a distance of 10.3$\pm$0.1~pc. In the Hipparcos Catalogue
Pollux was flagged as a possible micro-variable with a photometric amplitude 
of less than 0.03~mag (but no obvious periodicity), as well as a possibly 
non-single star, maybe because of slightly different astrometric solutions 
from the two different data reduction consortia. 

\citet{walker89} were the first to report significant radial velocity (RV)
variations for Pollux, with a standard deviation of 26~m/s around the mean
from RV measurements spread over about five years. Though they noted that 
based on a periodogram analysis significant periodicity was present in the data,
they did not quote any period. Only after having monitored Pollux extensively
over twelve years with a typical RV precision of 10--20~m/s, \citet{larson93}
published a RV period of 584.65$\pm$3.3~days and discussed possible reasons for
the observed periodicity. Possible explanations include an orbiting planetary
companion or rotational modulation of surface features. The latter hypothesis 
was supported by a slight indication in the data for a periodicity in the
equivalent width index data of the 8662{\AA} (Ca~{\sc ii}) line with about the 
same period as found in the radial velocities, but with a very low amplitude 
and only a marginal statistical significance.

Finally, \citet{hatzes93} presented again strong evidence for a periodicity
in the radial velocities with a period of 558~days. The spectra were taken
over a period of 3.5~years, and the typical RV accuracy was 20~m/s.
The RV variations were consistent in amplitude and phase with the older 
data by \citet{walker89}.

Here we present again precise radial velocity measurements of Pollux which
leave no doubt about a periodicity, determined from our data to 589.7~days.
This RV set is the most comprehensive and precise one taken so far for this star,
spanning almost six years. From the first measurements of Walker in 1981 to
the latest ones by us in 2006, this adds up to 25~years of RV monitoring for
Pollux, with no evidence for a change in phase or amplitude of the almost
sinusoidal variations. Along with no detectable variations in the
spectral line shapes, our data set thus lends further evidence for the
companion hypothesis.

In Section~\ref{obs}, we describe our observations which are part of a larger
program of monitoring giant stars for periodic RV changes and present our
orbital fit to the RV data. In Section~\ref{ana}, we analyze the spectral
line shapes with the help of the bisector method, and in Section~\ref{disc}
we present a discussion and our final conclusions.

\section{Observations}
\label{obs}

\begin{deluxetable}{lc}
\tablewidth{0pt}
\tablecolumns{2}
\tablecaption{\label{orbit} Fitted and derived orbital parameters}
\tablehead{\multicolumn{1}{l}{Parameter} & \colhead{Value}}
\startdata
Period [days]                        & $589.7  \pm 3.5$ \\
$T_0$ [JD-2\,450\,000]               & $2337.9^{+70}_{-52}$ \\
Eccentricity                         & $0.06 \pm 0.04$ \\
$\omega$ [deg]                       & $277 \pm 8$   \\
$f(m)\ [10^{-9}\rm{M_{\odot}}]$      & $6.2 \pm 0.6$ \\
$m_2\sin i$ [\mjup]\tablenotemark{a} & $2.9 \pm 0.1$ \\
Semi-major axis [AU]                 & $1.69 \pm 0.03$ \\
RV semi-amplitude [\ms]              & $46.9 \pm 1.5$ \\
Reduced $\chi^2$                     & 6.3 \\
rms scatter around fit [\ms]         & 9.0 \\
\enddata

\tablenotetext{a}{The companion mass error does not include the uncertainty
in the stellar mass.}

\end{deluxetable}

The observations were carried out as part of a larger program 
measuring precise radial velocities of several hundred G and K~giant
stars at Lick Observatory. The early objectives of this program have been
described in \citet{frink01}, and the first substellar companion from the
survey (around the K~giant $\iota$~Dra) was announced in \citet{frink02}.
\citet{hekker06} characterize the giant stars with the most stable radial 
velocities in our survey, and 
\citet{mitchell07} present evidence for four more K~giant stars harboring
one or more substellar companions.

\begin{figure}
\plotone{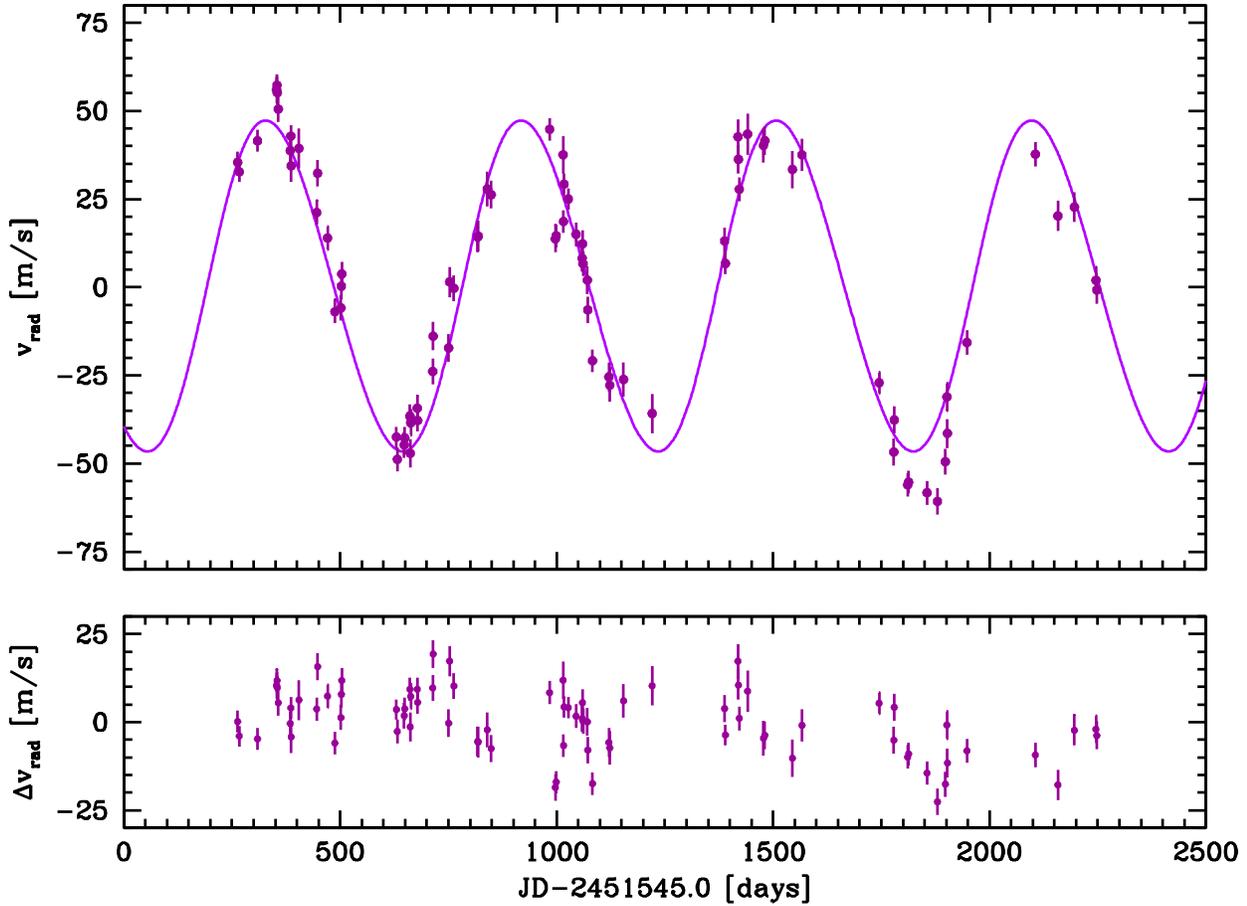}
\caption{\label{vrad} Top Panel: Radial velocities measured at Lick Observatory,
along with error bars, covering about 5.5~years from September~2000 to
February~2006. The best fit Keplerian is overplotted, with a period of
589.7~days. Bottom Panel: Radial velocity residuals after the best fit 
Keplerian has been subtracted. The remaining radial velocity scatter has
a standard deviation of 9~m/s, and no systematics are visible in the
residuals, neither by eye nor in the periodogram of the residuals in the lower
part of Fig.~\ref{losca_vrad}.}
\end{figure}

As part of this ongoing program, we have obtained 80~spectra for Pollux
covering about 5.5~years. All observations were taken using the Hamilton
High Resolution Echelle Spectrograph (R$\approx$60\,000 at 6000~\AA)
at Lick Observatory, attached to the 0.6~m Coud\'e Auxiliary Telescope (CAT).
Typical exposure times were 1.5~minutes for Pollux, yielding a S/N of
up to 200. However, some observations were taken with cloud cover, where
exposure times can be considerably longer and the S/N somewhat smaller
(around 150). The individual radial velocities, obtained as described
in \citet{butler96}, are listed in Table~\ref{rvobs},
along with their formal errors. Figure~\ref{vrad} shows the measurements
together with a Keplerian fit to the data; the orbital elements are listed
in Table~\ref{orbit}. There is no doubt about the clear periodicity in the
data, as has already been observed by \citet{walker89}, \citet{larson93} and
\citet{hatzes93}. 

\begin{figure}
\plotone{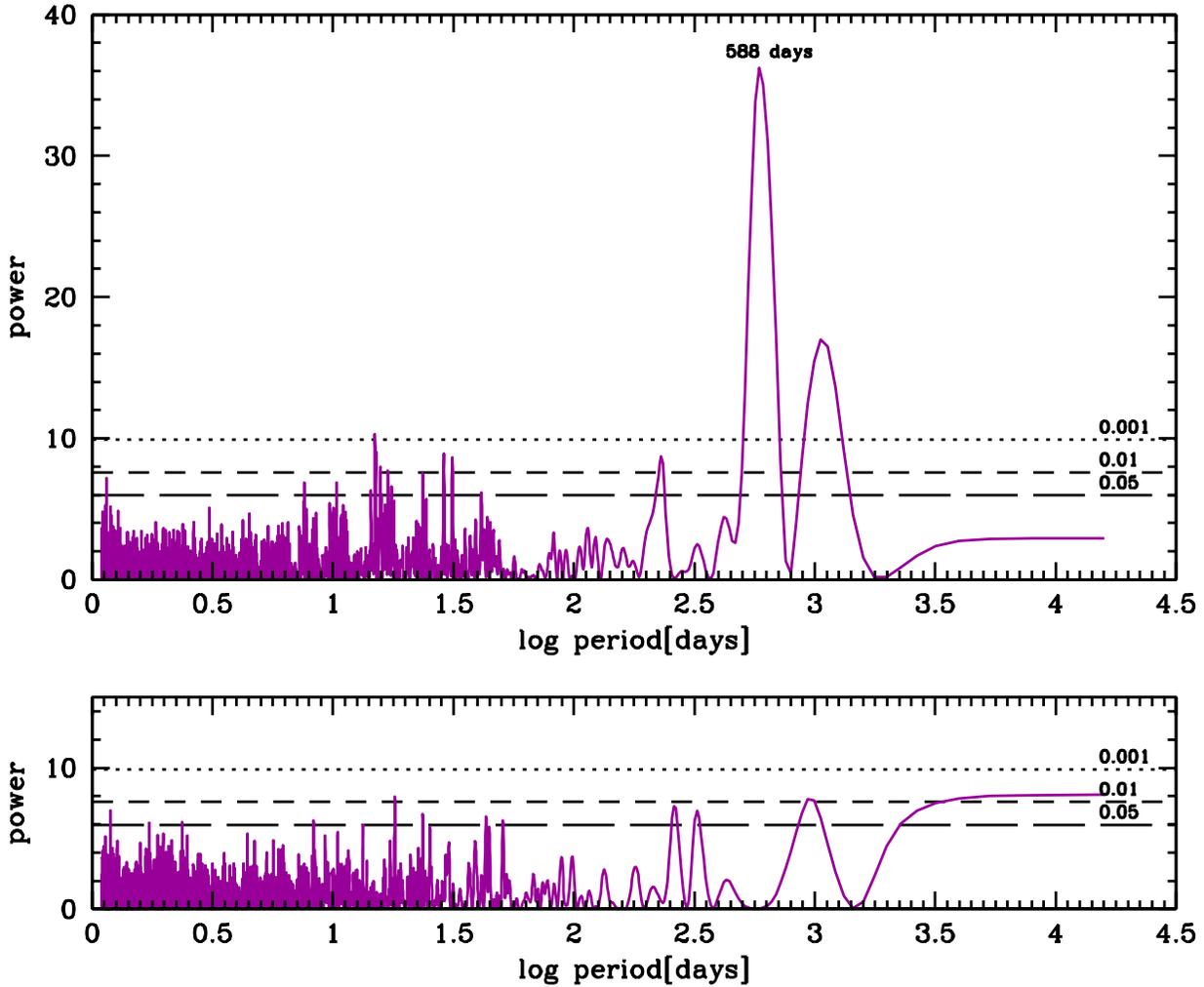}
\caption{\label{losca_vrad} Top Panel: periodogram of the measured radial
velocities. The highly significant peak occurs at 588~days; a Kepler fit to
the data reveals a best fit period of 589.7~days. The next significant peak
to the right, at about log P[days]=3.07, corresponds to twice the value
of the most significant period. The numbers at the right indicate the false alarm
probabilities of the labeled lines; a highly significant peak in the periodogram
would sit clearly above the highest line, indicating a false alarm probability of
less than 0.1\%.
Bottom panel: same as above, but with the Kepler fit
corresponding to 589.7~days removed from the radial velocities. No significant
peak is left in the periodogram of the radial velocity residuals.}
\end{figure}

Nevertheless, there is some additional scatter at a level of 9~m/s present in
the data, which is larger than expected based on the formal measurement
errors. It is
possible that this stems from radial pulsations, but with the theoretical
period for the fundamental mode shorter than 1~day \citep{hatzes93} our
sampling is inadequate to provide any further constraints. 
Solar-like oscillations in late G~giants have been found by \citet{frandsen02}
and \citet{deridder06}, and it is not unreasonable to assume that similar
oscillations in Pollux are responsible for the excess jitter. 
The amplitude however is not large enough to affect the derivation of the
orbital parameters of the putative companion.
Furthermore, there are no indications for any additional periodicities, as can
be seen in the lower part of Fig.~\ref{vrad} where the orbital fit has been
subtracted from the data, and the corresponding Lomb-Scargle periodogram 
in the lower part of Fig.~\ref{losca_vrad}. 

\begin{figure}
\plotone{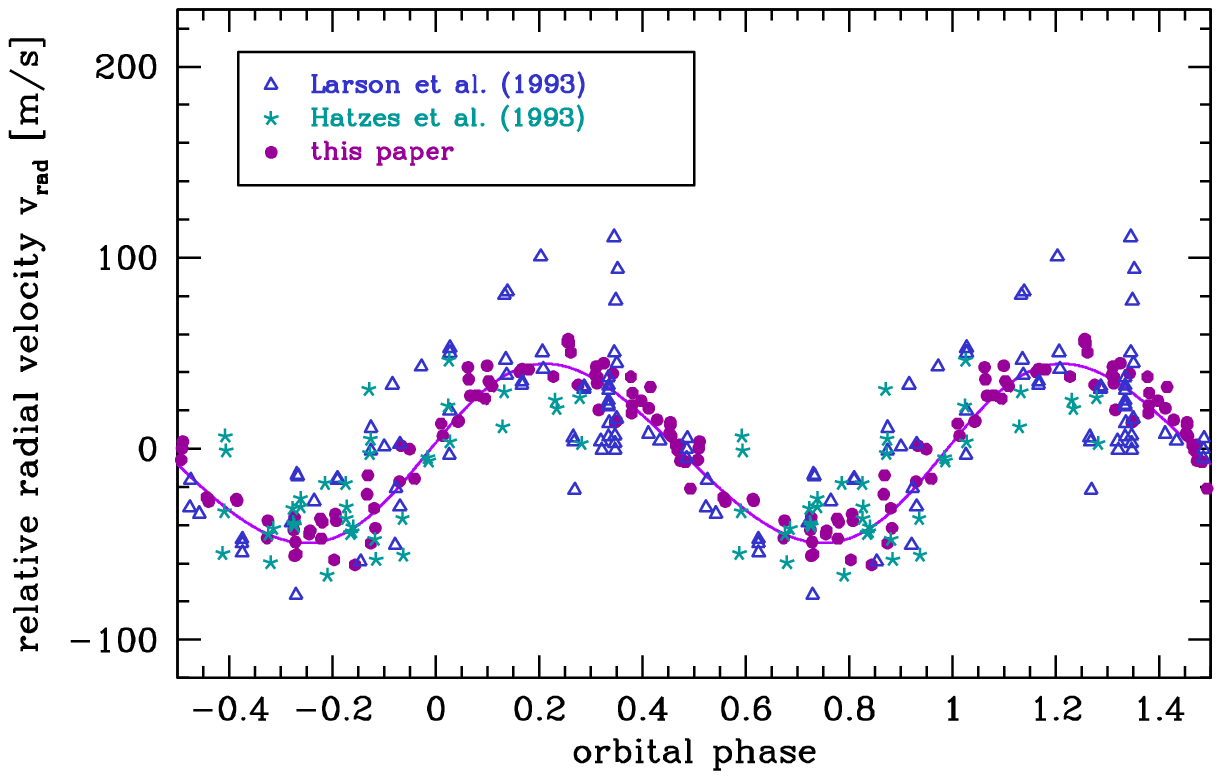}
\caption{\label{phased_vrad} Radial velocities from \citet{larson93}
(open triangles) and \citet{hatzes93} (asterisks) phased to the period determined from our 
own measurements (filled circles). Our orbital fit is also shown (solid line). 
No changes in period or amplitude of the periodic RV signal is apparent from
that plot. For clarity, no individual error bars have been included in the plot.
The data in \citet{larson93} consist of two separate datasets; data from CFHT have 
mean internal errors of 12~m/s, data from DAO 27~m/s. The majority of the datapoints
in \citet{hatzes93} has mean internal errors of around 20~m/s, while a few datapoints
have mean internal errors of 7~m/s. Note that not all of the measurements in \citet{hatzes93} 
have been used; the second set of 13 measurements in their Table~1C is identical to the
first 13~measurements and thus obviously wrong, so it has been omitted. 
Errors for the DAO dataset
from \citet{larson93} are rather inhomogeneous and reach up to around 80~m/s for some
measurements, which explains the few data points which seem to be outliers in 
this graph.}
\end{figure}

Our period for the orbital fit is 589.7$\pm$3.5~days and
compares very well with the value of \citet{larson93} (584.65$\pm$3.3~days),
while it is somewhat larger than the period derived by \citet{hatzes93}
(554$\pm$8~days). All previous measurements are consistent in amplitude and 
phase with our result; Fig.~\ref{phased_vrad} shows a phased plot of our measurements
along with the earlier ones by \citet{larson93} and \citet{hatzes93}.
(Note that we measure only relative radial velocities with an arbitrary zero
point, so that a vertical shift had to be applied before plotting them along
with the other data sets. This vertical shift is a free parameter
in our fit; its formal error is 1.1~m/s and should be negligible for the
comparison.) A combined fit to all RV data produces very similar orbital
parameters as a fit to our RV data alone; the period from the combined fit is
larger by 1.6~days as compared to the period quoted in Table~\ref{orbit} 
based on our RV data alone, very well within the formal error.

The minimum companion mass derived from our orbital fit is 2.9~\mjup\ assuming
a stellar mass for Pollux of 1.86~\msun. The stellar mass was derived from the
location of the star in the color-magnitude diagram as determined from
Hipparcos data and compared to the evolutionary tracks from
\citet{girardi2000}. Solar metallicity was assumed for the comparison, which is 
a good approximation \citep{drake91}. \citet{allende99} derive a value of
1.7~\msun\ for the mass of Pollux with a very similar method as described
above, and \citet{drake91} also derive a mass of 1.7~\msun\ from a detailed
model atmosphere analysis. Using this value for the stellar mass would yield
a minimum mass of 2.7~\mjup. The error on the minimum companion mass due to
the error in the knowledge of the primary mass is thus about 0.2~\mjup, so that
the total error (including the formal error derived from the orbital fit, see
Table~\ref{orbit}) amounts to about 0.3~\mjup.

Together, the radial velocities cover about 25~years, and the variations have
been rather consistent over that time. Nevertheless, though we consider the
interpretation of the observed RV changes as the result of an orbiting
companion likely, it is possible that another mechanism might cause the
observed RV variations. In Chapter~\ref{ana}, we take a closer look
at the spectral line shapes which might provide further hints at the 
underlying mechanism.

\onecolumn

\twocolumn

\section{Line Shape Analysis}
\label{ana}

In order to investigate whether the observed RV variations are caused by a shift
of the spectral lines as a whole (as expected in the presence of a companion)
or by a change in the symmetry of the spectral lines giving rise to a net
change in RV (as expected in the presence of pulsations), bisectors of the
cross correlation profile have been analyzed. We used all spectral lines
between about 6540~{\AA} and 9590~{\AA} from 29~spectral orders and obtained an
average line profile by correlation with a synthetic template which was obtained
from the VALD database 
\citep[][available at \url{http://ams.astro.univie.ac.at/vald/}]{kupka99},
matching the effective temperature and surface gravity of Pollux. The spectral
range from about 5000~{\AA} to 5800~{\AA} could not be used because it is
affected by iodine lines, and the spectral range around 6300~{\AA} was avoided 
because it is dominated by strong telluric oxygen lines (which give rise to a
spurious one year period when included in the analysis). Otherwise, as many
lines as possible from a continuous spectral range were used, since it is known
\citep{gray83,gray84} that different lines in the same star can display
different bisector behavior. 

For the cross correlation with the synthetic template, our individual spectra
had to be wave\-length\--cali\-bra\-ted, for which we used the Thorium-Argon exposure
which was taken closest in time to each spectrum, either from the
beginning or end of the night. Altogether, about 1110 spectral lines with 
theoretical depths between 0.1 and 0.9 were used for the cross correlation.
Of course many blended lines are included in the cross correlation, but by
using many lines we are confident that the effects average out. Furthermore, as
long as the same lines are used for all observations, and only variations in
the shape are of interest, blends by stellar lines do not affect the final result.

After having obtained the cross correlation profiles, we determined bisectors
by stepping down the blue side of the profile, linearly interpolating the line
depth at the red side for the same flux level as observed at the blue side
(in the center of the profile, a parabolic fit was used instead of the linear
one), and derived the midpoints between the velocities on the blue and red side
of the profile. The connection of midpoints determined in this way is the
bisector; see \citet{povich2001} for more details on the method.

In order to analyze possible variations in the bisector, two quantities are 
defined which characterize its shape: the velocity span and the velocity displacement. 
While the bisector span is the difference between the
width of the bisector at two different flux levels (30\% and 75\% were used
here), the velocity displacement is the average width of the bisector at three
different flux levels (30\%, 60\% and 75\% were used). 

\begin{figure}
\plotone{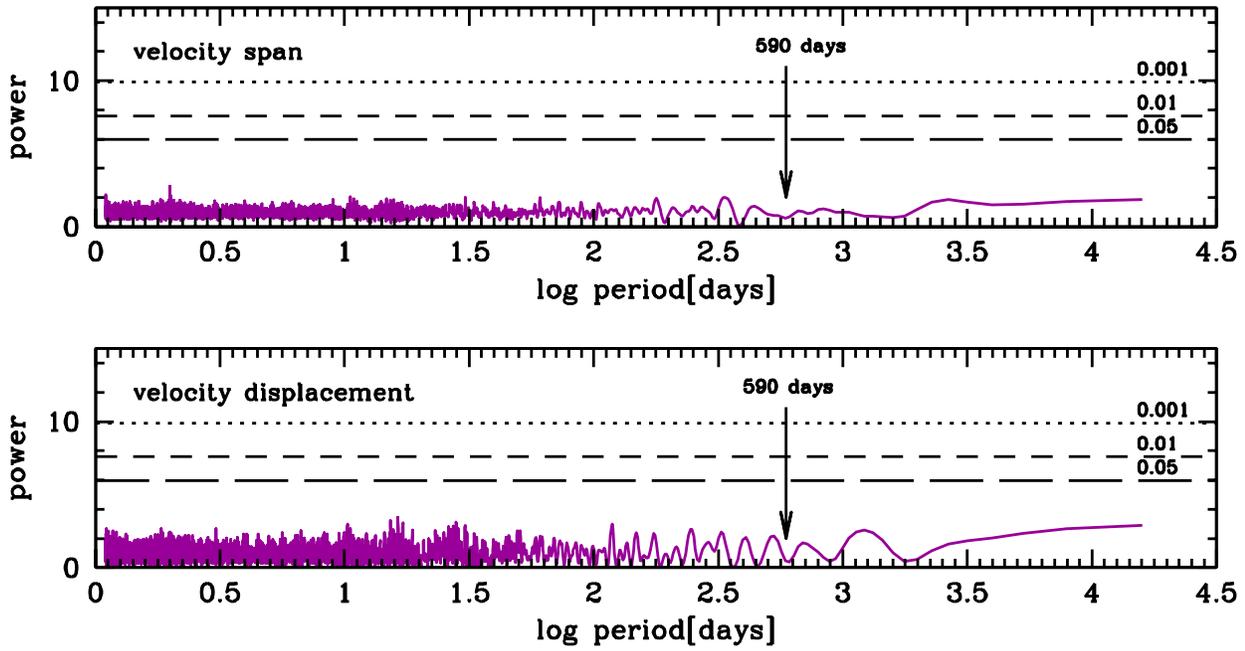}
\caption{\label{losca_bisect} Periodograms of the derived bisectors. Top panel:
velocity span, bottom panel: velocity displacement. There are no significant
periodicities visible; all trial periods have an extremely small probability of 
being real. The period present in the RV data is indicated.}
\end{figure}

Periodograms of both quantities are shown in Fig.~\ref{losca_bisect}.
No significant periods whatsoever are present in these periodograms; all trial
periods have extremely small significance levels. In particular, no peak is
present at the RV period, so that we conclude that there is no evidence for any
variations in the shapes of the spectral lines in our Pollux spectra. 

This finding is consistent with the one by \citet{hatzes98}, who also analyzed
the bisectors of two different spectral lines in their Pollux data without 
discovering any periodicities similar to the RV period (they used 554~days). 
However, they caution that low-order non-radial pulsation modes, which might be
able to account for the observed RV variations, might produce changes in the
bisector velocity span of only 5 to 20 m/s. While we estimate the errors
in a single bisector velocity span to lie between 50 and 100~m/s in our analysis
(50~m/s in the analysis of \citet{hatzes98}), the sensitivity to a real periodic 
signal in the data increases if one uses a large number of spectra for the periodogram,
to about the relevant level. However, we conclude that
non-radial pulsation cannot be completely ruled out with the
observational material or analysis methods currently available.

\onecolumn

\twocolumn

\section{Conclusions}
\label{disc}

The presently available data on Pollux are all compatible with an orbiting
substellar companion around this bright and nearby star. The minimum mass of
the companion is 2.9$\pm$0.3~\mjup, which makes it most likely a planet and not a
brown dwarf. From the non-detection of the companion in the Hipparcos data, one can 
derive an upper limit on the companion mass, since it would have been detected by
Hipparcos if it had been massive enough. 
The lower limit on the inclination which we derive from Hipparcos is
about 5\degr, which translates into an upper mass limit of 33~\mjup,
constraining the companion to be of substellar nature (if the companion
interpretation of the RV pattern is correct).
Its period is 589.7$\pm$3.5~days, and it orbits at a distance
of 1.69$\pm$0.03~AU from the star in an almost circular orbit. The RV pattern
has been stable over the last 25~years. An analysis of the spectral line shapes
shows no evidence for any changes with the RV period nor any other periods,
which one might expect in the presence of pulsations. Hipparcos has
picked up excess scatter in the astrometric standard solution (without
companion), but it is unlikely that this is the signature of the planetary
companion, since its expected minimum astrometric signature is only
50~\muas.

Detailed theoretical predictions of the expected amount of bisector asymmetry
in the presence of non-radial g- or r-mode pulsations in giant stars are
currently not available. However, the numerical simulations conducted
by \citet{hatzes96} show that it is in principle possible to explain the observed 
RV variations in Pollux by low-order non-radial pulsation modes, while the
accompanying bisector variations would be too small to be detected with current
techniques \citep{hatzes98}.
This is the main reason why some last doubt remains about the
interpretation of the RV variations in terms of an orbiting companion, even
if it would be a bit surprising to find only one single long-period 
pulsation mode.
Note that the small eccentricity which we find (0.06$\pm$0.04) is barely
significantly different from zero, so that it cannot be used to rule out
pulsations as the reason for the observed RV variations as has been done for
highly eccentric giant star orbits \citep{frink02}.

In contrast to that, rotational modulation of starspots can be excluded as the
reason for the observed RV changes, since otherwise some photometric
variability larger than the micro-variability actually seen should have been
detected by Hipparcos. Also, it
would be difficult to explain how a single or several starspots could produce
RV variations which are so close to sinusoidal over the rotation period as well
as stable over the last 25~years.

We conclude that while evidence is accumulating and compelling for an orbiting
planet around Pollux, the final confirmation has to await a theoretical
prediction of the amount of spectral line asymmetry in the presence of
non-radial g- or r-mode pulsations in giant stars, much increased sensitivity 
in bisector analyses or photometry or the detection of the companion with independent 
techniques like e.g.\ precise astrometry.

After this paper was first submitted we learned of the similar paper by 
\citet{hatzes06}. \citet{hatzes06} present 55 new radial velocity measurements of 
Pollux with mean internal errors between 11 and 17~m/s taken between 1998 and 2006 
and analyze them together with the older datasets by \citet{larson93} and \citet{hatzes93}
also used in the present paper. Their orbital elements are in excellent agreement with
the ones derived in the present paper, and both papers arrive at the same 
conclusions regarding the interpretation of the observed RV periodicity in terms
of an orbiting planetary companion.

\acknowledgments

We kindly thank the staff at Lick Observatory for their outstanding
dedication and support, as well as our referee, Gordon Walker, for
helpful comments on the manuscript.

This research has made use of the 
Vienna Atomic Line database (VALD) located at\\
\url{http://ams.astro.univie.ac.at/vald/}.

\begin{deluxetable}{crc}
\tablecaption{\label{rvobs} Measured radial velocities for Pollux}
\tabletypesize{\scriptsize} \tablewidth{0pt} \tablehead{
\colhead{JD\tablenotemark{a} [days]} & \colhead{$v_{\mbox{\tiny rad}}$ [\ms]} &
\colhead{$\sigma_{v_{\mbox{\tiny rad}}}$ [\ms]}}
\startdata
 1808.039 & $   35.4 $ &  3.1 \\
 1812.039 & $   32.7 $ &  2.9 \\
 1854.073 & $   41.5 $ &  3.1 \\
 1898.060 & $   56.0 $ &  3.8 \\
 1898.940 & $   57.3 $ &  3.1 \\
 1899.899 & $   55.1 $ &  3.2 \\
 1901.898 & $   50.5 $ &  3.7 \\
 1929.819 & $   38.7 $ &  3.5 \\
 1930.893 & $   42.9 $ &  3.2 \\
 1931.723 & $   34.4 $ &  4.5 \\
 1949.771 & $   39.4 $ &  5.7 \\
 1990.667 & $   21.2 $ &  3.3 \\
 1992.693 & $   32.3 $ &  3.8 \\
 2015.697 & $   13.9 $ &  3.2 \\
 2032.681 & $   -6.9 $ &  3.3 \\
 2046.656 & $   -5.9 $ &  3.6 \\
 2047.657 & $    0.3 $ &  3.7 \\
 2048.659 & $    3.8 $ &  3.5 \\
 2175.055 & $  -42.5 $ &  2.9 \\
 2177.047 & $  -48.8 $ &  3.3 \\
 2193.021 & $  -44.7 $ &  3.6 \\
 2193.983 & $  -42.8 $ &  3.2 \\
 2205.968 & $  -36.6 $ &  3.3 \\
 2206.970 & $  -47.1 $ &  4.1 \\
 2207.995 & $  -38.4 $ &  3.5 \\
 2222.917 & $  -34.3 $ &  3.4 \\
 2223.924 & $  -37.8 $ &  3.1 \\
 2258.832 & $  -23.9 $ &  3.7 \\
 2259.898 & $  -13.9 $ &  4.0 \\
 2295.812 & $  -17.2 $ &  3.9 \\
 2297.873 & $    1.5 $ &  4.2 \\
 2307.749 & $   -0.3 $ &  3.7 \\
 2362.729 & $   14.2 $ &  4.3 \\
 2363.740 & $   14.5 $ &  4.4 \\
 2384.701 & $   27.9 $ &  4.9 \\
 2393.676 & $   26.2 $ &  3.9 \\
 2529.043 & $   44.8 $ &  3.3 \\
 2542.035 & $   13.7 $ &  3.8 \\
 2544.004 & $   14.6 $ &  3.1 \\
 2560.031 & $   37.6 $ &  5.4 \\
 2561.051 & $   18.6 $ &  3.2 \\
 2562.023 & $   29.2 $ &  3.1 \\
 2572.020 & $   25.0 $ &  3.1 \\
 2590.000 & $   15.0 $ &  3.4 \\
 2603.974 & $    8.2 $ &  3.7 \\
 2604.957 & $   12.3 $ &  3.8 \\
 2605.893 & $    6.7 $ &  3.5 \\
 2615.882 & $    2.0 $ &  4.0 \\
 2616.889 & $   -6.4 $ &  3.7 \\
 2627.889 & $  -20.9 $ &  3.2 \\
 2665.800 & $  -25.5 $ &  4.1 \\
 2667.840 & $  -27.8 $ &  4.7 \\
 2699.669 & $  -26.2 $ &  4.8 \\
 2765.687 & $  -35.8 $ &  5.6 \\
 2933.021 & $   13.1 $ &  3.8 \\
 2935.010 & $    6.7 $ &  3.0 \\
 2964.043 & $   42.7 $ &  4.9 \\
 2964.938 & $   36.3 $ &  4.0 \\
 2966.919 & $   27.8 $ &  3.4 \\
 2985.912 & $   43.4 $ &  5.8 \\
 3022.829 & $   40.2 $ &  4.9 \\
 3025.893 & $   41.6 $ &  4.0 \\
 3089.732 & $   33.4 $ &  5.3 \\
 3111.687 & $   37.5 $ &  4.5 \\
 3290.027 & $  -27.1 $ &  3.0 \\
 3323.988 & $  -46.7 $ &  3.8 \\
 3324.988 & $  -37.6 $ &  3.9 \\
 3355.851 & $  -56.1 $ &  3.3 \\
 3357.988 & $  -55.3 $ &  3.2 \\
 3400.828 & $  -58.3 $ &  3.3 \\
 3424.734 & $  -60.8 $ &  3.7 \\
 3442.703 & $  -49.5 $ &  3.6 \\
 3446.642 & $  -31.1 $ &  4.0 \\
 3447.665 & $  -41.5 $ &  4.1 \\
 3492.671 & $  -15.7 $ &  3.4 \\
 3651.058 & $   37.7 $ &  3.5 \\
 3703.043 & $   20.2 $ &  4.3 \\
 3741.009 & $   22.7 $ &  4.3 \\
 3790.775 & $    1.9 $ &  3.7 \\
 3792.756 & $   -0.8 $ &  3.9 \\
\enddata
\tablenotetext{a}{Julian date - 2\,450\,000}
\end{deluxetable}

\end{document}